\documentclass[journal]{IEEEtran}

\usepackage{tabularx} % in the preamble
\pdfoutput=1

\usepackage{cite}

\usepackage{hyperref}       % hyperlinks
\usepackage{url}            % simple URL typesetting
\usepackage{booktabs}       % professional-quality tables
\usepackage{amsfonts}       % blackboard math symbols
\usepackage{nicefrac}       % compact symbols for 1/2, etc.
\usepackage{microtype}      % microtypography
\usepackage{adjustbox}
\usepackage{hyperref}
\usepackage{bm}
\usepackage[caption=false,font=normalsize, labelfont=sf, textfont=sf]{subfig}
\usepackage[english]{babel}

\usepackage{multirow, rotating, array}
\usepackage{amsfonts, amssymb, amsmath}

\usepackage{floatrow}
\usepackage{flafter}

\usepackage{color, float}
\usepackage{afterpage}
\usepackage{flafter}

\usepackage{algorithmic}
\usepackage[algoruled]{algorithm2e}

\usepackage{wrapfig}
\usepackage{makeidx}
\usepackage{makecell}

\hypersetup{
	colorlinks   = true, %
	urlcolor     = blue, %
	linkcolor    = red, %
	citecolor    = blue   %
}

\newcommand{\add}[1]{\textcolor{magenta}{#1}}

\renewcommand{\add}[1]{#1}

\newcommand{\addminor}[1]{\textcolor{magenta}{#1}}

\renewcommand{\addminor}[1]{#1}

\IEEEpubid{\textbf{Accepted in IEEE Transactions on Medical Imaging, 2019}}

\IEEEpubid{\begin{minipage}{\textwidth}\ \\[12pt]
		 \\ 
		Accepted in IEEE Transactions on Medical Imaging, 2019
\end{minipage}}

\begin{document}

\bstctlcite{IEEEexample:BSTcontrol}

\title{\add{Deep learning analysis of coronary arteries in cardiac CT angiography for detection of patients requiring invasive coronary angiography}}

\author{
	Majd~Zreik, Robbert~W.~van~Hamersvelt, Nadieh~Khalili, Jelmer~M.~Wolterink, \\Michiel Voskuil, Max~A.~Viergever, Tim~Leiner, Ivana~I\v{s}gum%
	\thanks{M.~Zreik and N.~Khalili are with the Image Sciences Institute, University Medical Center Utrecht, The Netherlands (e-mail: m.zreik@umcutrecht.nl).}%
	\thanks{R.~W.~van~Hamersvelt is with the Department of Radiology, University Medical Center Utrecht, The Netherlands.}%
	\thanks{J.~M.~Wolterink is with the Image Sciences Institute, University Medical Center Utrecht, The Netherlands and the Department of Biomedical Engineering and Physics, Amsterdam University Medical Center.}	
	\thanks{M.~Voskuil is with the Department of Cardiology, University Medical Center Utrecht and Utrecht University, The Netherlands.}
	\thanks{M. A.~Viergever is with the Image Sciences Institute, University Medical Center Utrecht and Utrecht University, The Netherlands.}
	\thanks{T.~Leiner is with the Department of Radiology, University Medical Center Utrecht and Utrecht University, The Netherlands.}
	\thanks{I.~I\v{s}gum is with the Image Sciences Institute, University Medical Center Utrecht, The Netherlands, the Department of Biomedical Engineering and Physics, Amsterdam University Medical Center, and the Department of Radiology and Nuclear Medicine, Amsterdam University Medical Center.}
	\thanks{This study was financially supported by the project FSCAD, funded by the Netherlands Organization for Health Research and Development (ZonMw) in the framework of the research programme IMDI (Innovative Medical Devices Initiative); project 104003009.}
	\thanks{Copyright (c) 2019 IEEE. Personal use of this material is permitted. However, permission to use this material for any other purposes must be obtained from the IEEE by sending a request to pubs-permissions@ieee.org.}}
\markboth{}{}

\maketitle
	
%\footnote{corresponding author, \texttt{m.zreik@umcutrecht.nl}}

\begin{abstract}

In patients with obstructive coronary artery disease, the functional significance of a coronary artery stenosis needs to be determined to guide treatment. This is typically established through fractional flow reserve (FFR) measurement, performed during invasive coronary angiography (ICA). We present a method for automatic and non-invasive detection \add{of patients requiring ICA}, employing deep unsupervised analysis of complete coronary arteries in cardiac CT angiography (CCTA) images. 
We retrospectively collected CCTA scans of 187 patients, 137 of them underwent invasive FFR measurement in 192 different coronary arteries. These FFR measurements served as a reference standard for the functional significance of the coronary stenosis. The centerlines of the coronary arteries were extracted and used to reconstruct straightened multi-planar reformatted (MPR) volumes. 
To automatically identify arteries with functionally significant stenosis \add{that require ICA}, each MPR volume was encoded into a fixed number of encodings using two disjoint 3D and 1D convolutional autoencoders performing spatial and sequential encodings, respectively. Thereafter, these encodings were employed to classify arteries using a support vector machine classifier.
The detection of coronary arteries requiring invasive evaluation, evaluated using repeated cross-validation experiments, resulted in an area under the receiver operating characteristic curve of $0.81 \pm 0.02$ on the artery-level, and $0.87 \pm 0.02$ on the patient-level. 
\add{The results demonstrate the feasibility of automatic non-invasive detection of patients that require ICA and possibly subsequent coronary artery intervention.} This could potentially reduce the number of patients that unnecessarily undergo ICA.

\end{abstract}


%\linenumbers

\section{Introduction}

Obstructive coronary artery disease (CAD) is the most common type of cardiovascular disease \cite{benjamin2018heart}. Obstructive CAD develops when atherosclerotic plaque builds up in the wall of the coronary arteries, narrowing the coronary artery lumen \cite{cury2016cad}. This is defined as coronary stenosis, which can potentially limit blood supply to the myocardium, and could lead to ischemia and irreversible damage \cite{Pijl96}. Only functionally significant stenoses, i.e. those stenoses which significantly limit blood flow, need to be invasively treated in order to reduce CAD morbidity \cite{Pijl96,Toni09,Pijl10,Nune15}. Contrarily, invasively treating a functionally non-significant stenosis may lead to harmful output \cite{Pijl10,Pijl13}. Therefore, it is crucial to assess the functional significance of a coronary stenosis to guide treatment.

Cardiac CT angiography (CCTA) is typically used to noninvasively identify patients with suspected CAD and visually detect coronary artery stenosis \cite{Budo08a}. Although CCTA has high sensitivity 
in determining the functional significance of the stenosis, its specificity for this task is low \cite{Meij08, Bamb11,Ko12}. Therefore, to determine whether a coronary artery stenosis is functionally significant, patients with obstructive CAD typically undergo invasive coronary angiography (ICA) to measure the fractional flow reserve (FFR) in the coronary arteries. FFR is currently the reference standard for establishing the functional significance of a coronary stenosis and it is used to guide treatment \cite{Pijl96,Toni09}. However, because of the low specificity of CCTA, up to 50\% of patients undergo invasive FFR measurement unnecessarily \cite{Ko12}. To reduce the number of unnecessary invasive procedures, noninvasive determination of the functional significance of stenoses based on CCTA images has been intensively investigated. Several automatic methods for determination the functional significance of coronary artery stenosis in CCTA have been proposed \cite{zreik2018deep}. These methods can be divided into those that simulate and analyze blood flow in the coronary arteries \cite{Tayl13, Itu12, Nick15,itu2016machine}, and those that analyze and characterize the left ventricle (LV) myocardium \cite{zreik2018deep,Xion15}.

Methods that simulate and analyze the blood flow in the coronary arteries in CCTA images estimate FFR values along the coronary artery, which can be used to determine the functional significance of coronary artery stenosis. Taylor et al. \cite{Tayl13} were the first to propose noninvasive flow-based FFR estimation from CCTA images, \add{which was later validated in multiple clinical  studies\cite{Norg14,lu2017noninvasive}}. To determine FFR values along the coronary artery, computational fluid dynamics, coupled with assumptions of physiological boundary conditions, were used. 
Itu et al. \cite{Itu12} also presented a method to estimate FFR in the coronary artery tree in CCTA images by simulating blood flow. This method uses a parametric lumped heart model, while modeling the patient-specific hemodynamics in both healthy and diseased coronary arteries. Nickisch et al. \cite{Nick15} determined FFR values along the coronary artery by simulating blood flow and pressure along the coronary artery arteries using an electrical patient-specific parametric lumped model. 
Moreover, Itu et al. \cite{itu2016machine} presented a machine-learning-based model for estimating FFR along the coronary artery. The model is trained on a large number of synthetically generated coronary anatomies, where the target values are computed using a blood flow-based model \cite{Itu12}. \add{This method was further evaluated in \cite{coenen2018diagnostic}.}
While these techniques \cite{Tayl13,Itu12,Nick15,itu2016machine} achieved high accuracy, they are remarkably dependent on the accuracy of coronary artery lumen segmentation \cite{tesche2017coronary}. Manual annotation of the coronary artery lumen is a time consuming and a complex task, where commercially available automatic software tools typically require substantial manual interaction and correction, especially in CCTA scans with excessive atherosclerotic calcifications or imaging artefacts due to stents and cardiac motion \cite{Kiri13a}.

Recently, methods that do not model the blood flow in the coronary arteries but employ characteristics extracted from the myocardium in CCTA scans, have shown to be feasible. Our recent work \cite{zreik2018deep,van2018deep} presented a deep learning approach to automatically identify patients with a functionally significant coronary artery stenosis using analysis of the LV myocardium in CCTA. 
The method first characterizes the LV myocardium using a convolutional autoencoder (CAE). Thereafter, using the extracted characteristics, patients are classified according to the presence of functionally significant stenosis using an SVM classifier. 
Previously, Xiong et al. \cite{Xion15} presented a machine learning based approach to detect patients with anatomically significant stenosis using characteristics of the LV myocardium derived from a CCTA scan. In this method, the LV myocardium is aligned with the standard 17-segments model \cite{Cerq02a} to relate each myocardial segment to its perfusing coronary artery. Then, hand-crafted features, describing each myocardial segment, are extracted and used for supervised classification of patients according to the presence of anatomical significant stenosis. Thereafter, Han et al. \cite{Han17} employed the technique described in \cite{Xion15} to detect patients with functionally significant stenosis, as defined by the invasively measured FFR. Although these new methods \cite{zreik2018deep, Xion15} have presented promising results without the need for accurate coronary artery lumen segmentation, they still need to be validated in large and diverse patients cohorts. 

\add{Moreover, in our recent work \cite{zreik2018recurrent}, we have analyzed the coronary arteries employing a recurrent convolutional neural network (RCNN) for detecting and classifying the anatomical significance of the coronary artery stenosis. The RCNN employs a 3D convolutional neural network to extract local features along the coronary artery. Subsequently, a recurrent neural network aggregates the features to perform the classification tasks. However, such an approach cannot be directly employed for the detection of the functional significance of a coronary stenosis for two reasons. First, such RCNN only performs a local analysis of the artery, were the complete artery is not taken into account. Second, to train such an RCNN, local reference labels are required. Such a requirement is not practical in the case of the functional significance of a coronary stenosis, where FFR is used as the reference and is usually provided on the artery level only.}

\begin{figure}[t]
	\includegraphics[width=1\linewidth]{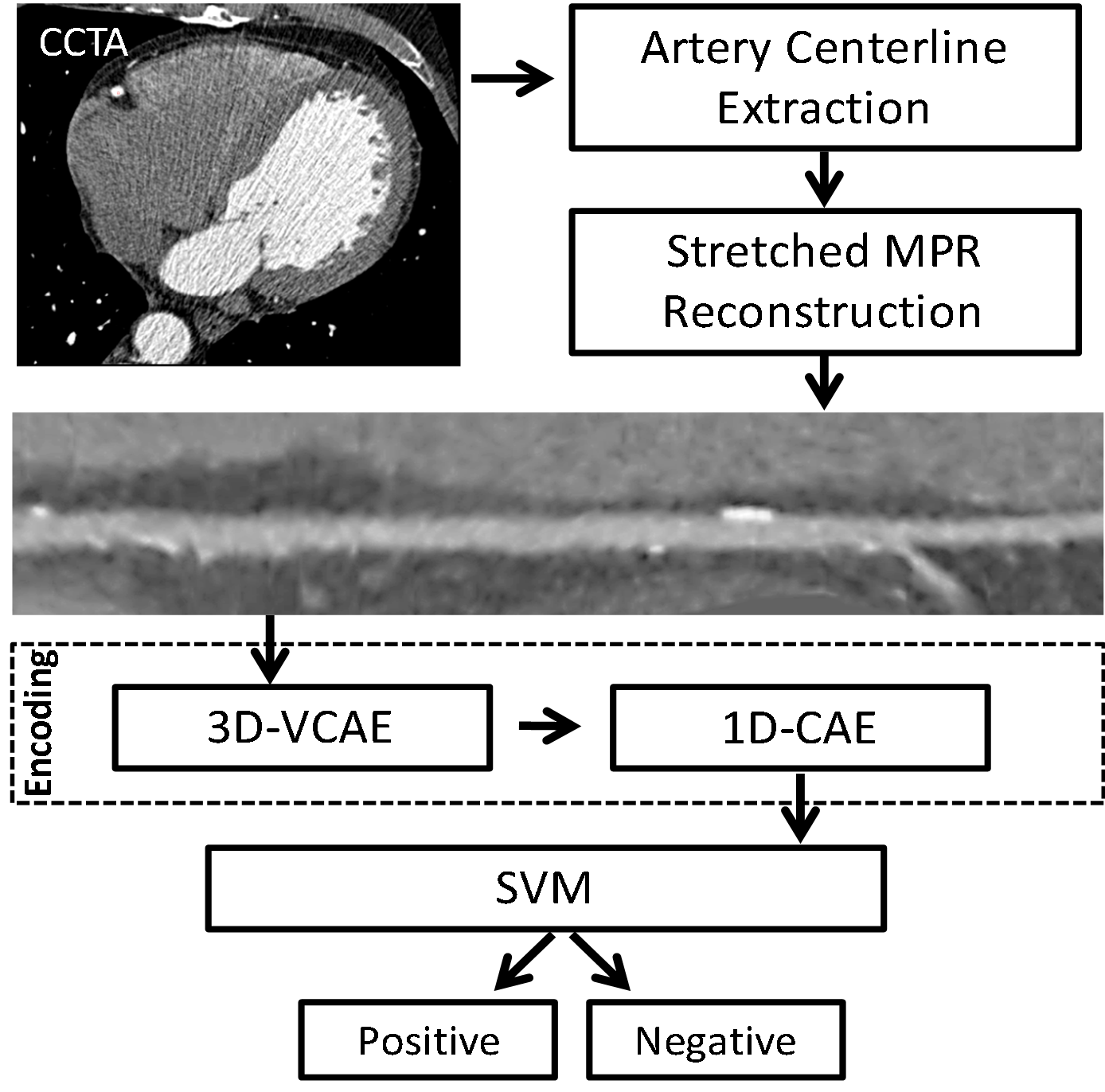}
	\caption{ Illustration of the proposed workflow. In a CCTA scan, the centerlines of the coronary arteries are extracted and used to reconstruct straightened multi-planar reformatted (MPR) images of the coronary arteries. Then, an unsupervised analysis is performed, where the MPR volume of a complete artery is encoded into a fixed number of encodings (features) using two disjoint convolutional autoencoders, applied sequentially: a 3D variational convolutional autoencoder (3D-VCAE), that spatially encodes local sub-volumes of the coronary artery, and a 1D convolutional autoencoder (1D-CAE), that sequentially encodes the encodings of the complete artery. Then the final extracted encodings are employed in a supervised fashion to classify arteries according to the \add{need of further invasive evaluation, in the form of ICA, to establish the need of intervention,} using a support vector machine (SVM) classifier.
	}

	\label{fig:flow}
	
\end{figure}

Here, we present a method to automatically and non-invasively identify coronary arteries \add{and patients requiring further invasive evaluation, i.e. ICA,} as determined by the invasively measured FFR. 
Blood flow in the coronary artery may be affected by multiple coronary artery stenoses and arterial plaques \cite{Pijl96,koo2011optimal}. Therefore, to \add{classify an artery according to the functional significance of the coronary artery stenosis}, local analysis of a single stenosis may be insufficient. Hence analysis of the complete artery should be performed.
Moreover, in clinical practice, usually a single, i.e. lowest, FFR value per coronary artery is reported.  
Consequently, employing supervised machine learning methods to directly analyze a whole volume of an artery (e.g. with 3D-CNN or \add{RCNN} \cite{zreik2018recurrent}) to detect the functional significance of each stenosis or estimating the invasively measured FFR values at every point along the coronary artery is unfeasible. 
Therefore, in the proposed work, a complete artery is analyzed in an unsupervised manner to extract lower-dimensional encoding, and thereafter to determine the presence of abnormal FFR.
First, using the extracted coronary artery centerline \cite{wolterink2019coronary}, the straightened 3D multi-planar reformatted (MPR) volume is reconstructed. Then, an MPR volume of a complete artery is characterized with a fixed number of encodings using convolutional autoencoders (CAEs) \cite{Masc11,Beng13,kingma2013auto}, which serve as unsupervised feature extractors.
As MPR volumes of complete coronary arteries have large volumetric sizes and variable lengths and shapes, a single traditional CAE cannot be successfully and directly applied to efficiently encode a complete artery. Therefore, in the here proposed work, two disjoint CAEs are employed. 
The first CAE performs spatial encoding of local sub-volumes along the artery. Then, a second CAE encodes the output of the first CAE - which depends on the artery length - into a fixed-length encoding. 
Finally, a support vector machine (SVM) \cite{Cort95} classifies arteries based on these encodings according to presence of functionally significant stenosis, as defined by the invasively measured FFR.
The proposed approach is illustrated in Fig.~\ref{fig:flow}. Our contributions are twofold. Firstly, we propose to jointly employ two disjoint CAEs that perform spatial and sequential encoding of large volumes with varying lengths. Secondly, in contrast to previous methods that detect the presence of functionally significant stenosis or determine FFR values non-invasively, our method does not require accurate and difficult to obtain segmentation of the coronary artery lumen or LV myocardium. Instead, it only requires the coronary artery centerline, which can be obtained automatically or semi-automatically \cite{Scha09a}.

The remainder of the manuscript is organized as follows. Section \ref{data} describes the data and reference standard. Section \ref{method} describes the method. Section \ref{results} reports our experimental results, which are then discussed in Section \ref{discussion}.

\section{Data} \label{data}

\subsection{Patient and Image Data}

This study includes retrospectively collected CCTA scans of 187 patients (age: $58.6 \pm 8.7$ years, 145 males) acquired between 2012 and 2016. The Institutional Ethical Review Board waived the need for informed consent. 

All CCTA scans were acquired using an ECG-triggered step-and-shoot protocol on a 256-detector row scanner (Philips Brilliance iCT, Philips Medical, Best, The Netherlands). A tube voltage of 120 kVp and tube current between 210 and 300 mAs were used. For patients $\le80$ kg contrast medium was injected using a flow rate of 6 mL/s for a total of 70 mL iopromide (Ultravist 300 mg I/mL, Bayer Healthcare, Berlin, Germany), followed by a 50 mL mixed contrast medium and saline (50:50) flush, and next a 30 mL saline flush. For patients $>80$ kg the flow rate was 6.7 mL/s and the volumes of the boluses were 80, 67 and 40 mL, respectively. Images were reconstructed to an in-plane resolution ranging from 0.38 to 0.56 mm, and 0.9 mm thick slices with 0.45 mm spacing.

In each CCTA scan, coronary arteries were tracked and their centerlines were extracted using the method previously described by Wolterink et al. \cite{wolterink2019coronary}. The method tracks the visible coronary arteries, where the arterial centerlines are extracted between the ostia and the most distal visible locations. Using the extracted centerlines, a 3D straightened MPR volumes with 0.3 $mm^3$ isotropic resolution were reconstructed for all coronary arteries and used for further analysis. \add{Note that we define an \textit{artery} as the vessel starting from the ostium until the most distal location visible in the CCTA.}

\subsection{FFR Measurements}

Out of the 187 patients, 137 patients suspected of obstructive CAD underwent invasive FFR measurements ($0.81 \pm 0.10$, interquartile range: 0.74-0.89), up to one year after the acquisition of the CCTA scan. In these patients, FFR was measured in 192 different arteries. FFR was recorded with a coronary pressure guidewire (Certus Pressure Wire, St. Jude Medical, St. Paul, Minnesota) at maximal hyperemia conditions. Maximal hyperemia was induced by administration of intravenous adenosine (at a rate of 140 $\mu$g/kg per minute) through a central vein. The FFR wire was placed at the most distal part possible in the target artery. Using manual pullback, a single minimal FFR value was assessed and recorded for each artery.

\section{Methods}\label{method}

Blood flow in the coronary artery may be affected by a single or multiple coronary artery stenoses \cite{Pijl96,koo2011optimal}\add{, located anywhere along the coronary artery; starting from the ostium until the most distal location visible in the CCTA}.
Therefore, to \add{classify an artery according to the functional significance of a stenosis}, local analysis of a single stenosis may be insufficient, \add{but the analysis of the complete artery is needed.}
Moreover, in clinical practice, determining invasive FFR values for each voxel within the artery lumen, or recording an FFR value for each point on the coronary artery centerline, is impractical and typically not performed. Instead, the minimal single FFR value per coronary artery is recorded, resulting in a single reference label per artery. 
\add{Hence, given the sparsity of reference labels along the artery, the large input dimensions and the limited dataset size, employing a supervised end-to-end machine learning methods (e.g. with 3D-CNN or RCNN \cite{zreik2018recurrent}) to directly detect the functional significance of each stenosis or estimating FFR values at every point along the coronary artery would be prone to overfitting.} 
Therefore, in the proposed work, an MPR of a complete artery is analyzed to determine the presence of abnormal FFR. First, to extract robust features of complete arteries, MPR volumes are characterized by a fixed number of encodings using convolutional autoencoders (CAEs) \cite{Masc11,Beng13,kingma2013auto}, regardless of the artery length. Then, the extracted encodings are used as input to an SVM classifier that determines whether the artery \add{needs further invasive evaluation, in the form of ICA, to establish the need of intervention.}

\subsection{Encoding the artery}\label{encoding_the_artery}

The main purpose of a convolutional autoencoder (CAE) is to extract robust compact features from unlabeled data, while removing input redundancies and preserving essential aspects of the data \cite{Masc11,Beng13,Lecu15}.
A CAE consists of two main parts, an encoder and a decoder \cite{Masc11,Beng13}. The encoder compresses the data to a lower dimensional latent space by convolutions and down-sampling. The decoder expands the compressed form to reconstruct the input data by deconvolutions and upsampling. 
A CAE is trained to minimize a difference loss between the encoder input and decoder output. This ensures that the encodings contain sufficient information to reconstruct inputs with low error \cite{Beng13}. Once the CAE is trained, the decoder is removed and the encoder is used to generate encodings for unseen data.

\begin{figure}[h]
	\includegraphics[width=1\linewidth]{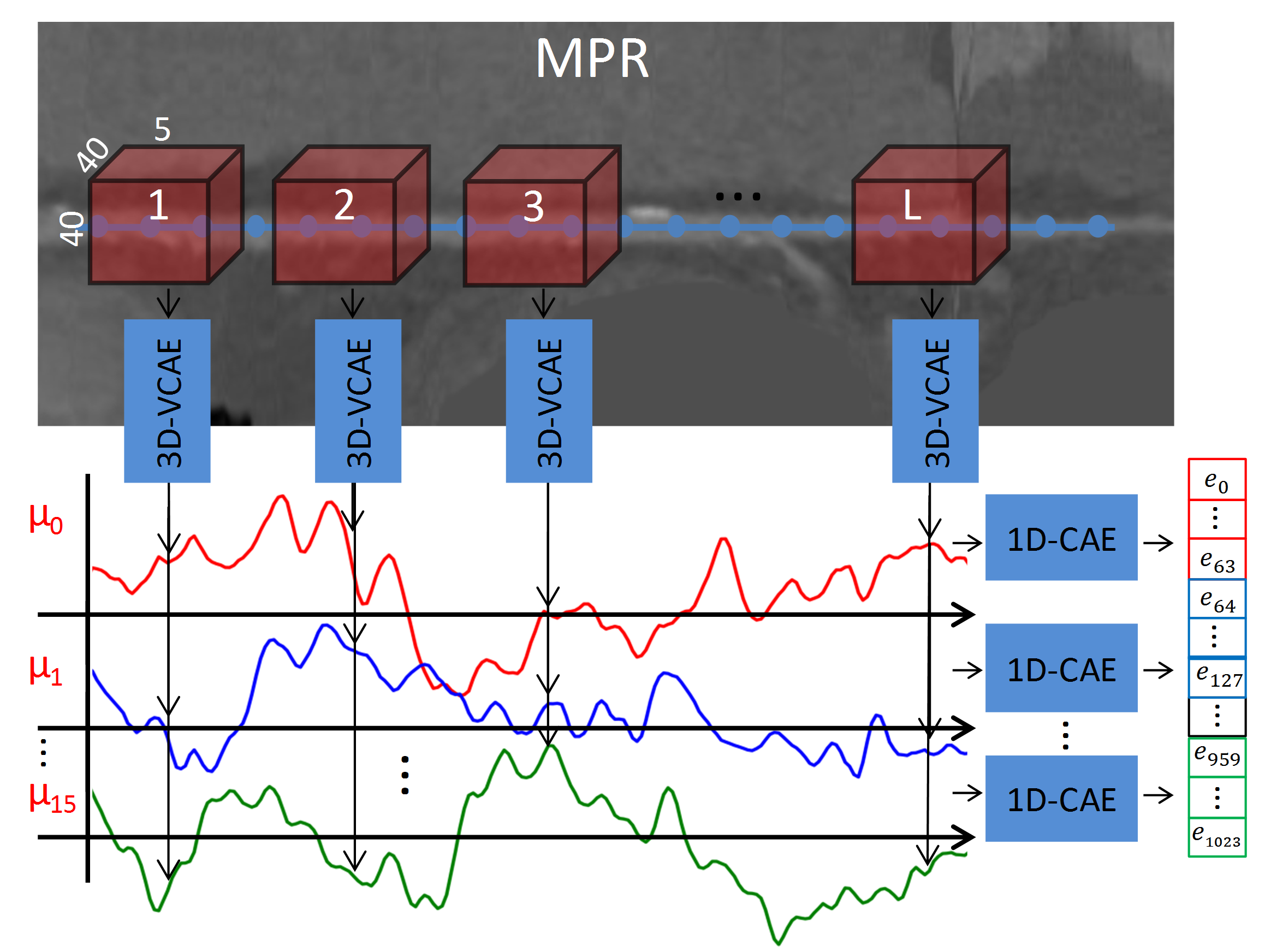}
	\caption{Illustration of the proposed encoding approach. To encode an MPR volume of a complete artery into a fixed number of encodings, a two stage encoding approach is applied. First, a 3D variational convolutional autoencoder (3D-VCAE) is applied to local 40x40x5 voxel sub-volumes extracted from the MPR along the artery. The 3D-VCAE encodes each volume into an encoding in a $R^{16}$ latent space. When applied to all sequential sub-volumes along the artery, the result is a feature map of the same height as the number of encodings and the same length as the artery length (L). This feature map is then represented as a set of individual 1D sequences of encodings. Each sequence contains an individual value in the $R^{16}$ latent space encoding, running along the artery.Then, a 1D convolutional autoencoder (1D-CAE) is applied separately to each of the 16 sequences of encodings and encodes each further into a second latent space with 64 dimensions ($R^{64}$). This results in a fixed number of encodings (1024) per artery, that represent the complete artery volume, regardless of its length and shape. 
		}
	\label{fig:encoding_flow}
	
\end{figure}

Coronary arteries are complex anatomical 3D structures, with varying lengths and anomalies across patients \cite{pannu2003current}. The resolution of modern CT scanners is high and a large number of voxels (millions) is contained in an MPR volume of a single artery. Therefore, following the straightforward approach of training a single CAE, applied directly to the complete artery volume without a large reconstruction error, is infeasible. Therefore, in this work, we propose a two-stage encoding approach to encode a complete MPR volume of the coronary artery, regardless of its length. Fig.~\ref{fig:encoding_flow} illustrates the proposed encoding flow. First, a 3D variational convolutional autoencoder (3D-VCAE) is applied to local sub-volumes extracted from the MPR along the artery centerline. As the 3D-VCAE is only applied to small input volumes, the number of its trainable parameters is relatively low.
The 3D-VCAE encodes each sub-volume into a set of small number of encodings. When applied to all sequential sub-volumes along the artery, the result is a \add{feature} map of the same height as the number of encodings and the same length as the artery length. This \add{feature} map is then represented as a set of individual 1D sequences of encodings. Each sequence contains an individual encoding out of the set of encodings, running along the artery (colored signals in Fig.~\ref{fig:encoding_flow}). This allows the analysis of complete arteries with varying length by a 1D convolutional autoencoder (1D-CAE), with low number of trainable parameters which decreases the chance of overfitting. Hence, the 1D-CAE encodes the varying length sequences of encodings further into a fixed number of encodings, that represent the complete artery, regardless of its length. 

\subsubsection{Spatial encoding with 3D variational convolutional autoencoder}

\begin{figure}[h]

	\includegraphics[width=\textwidth]{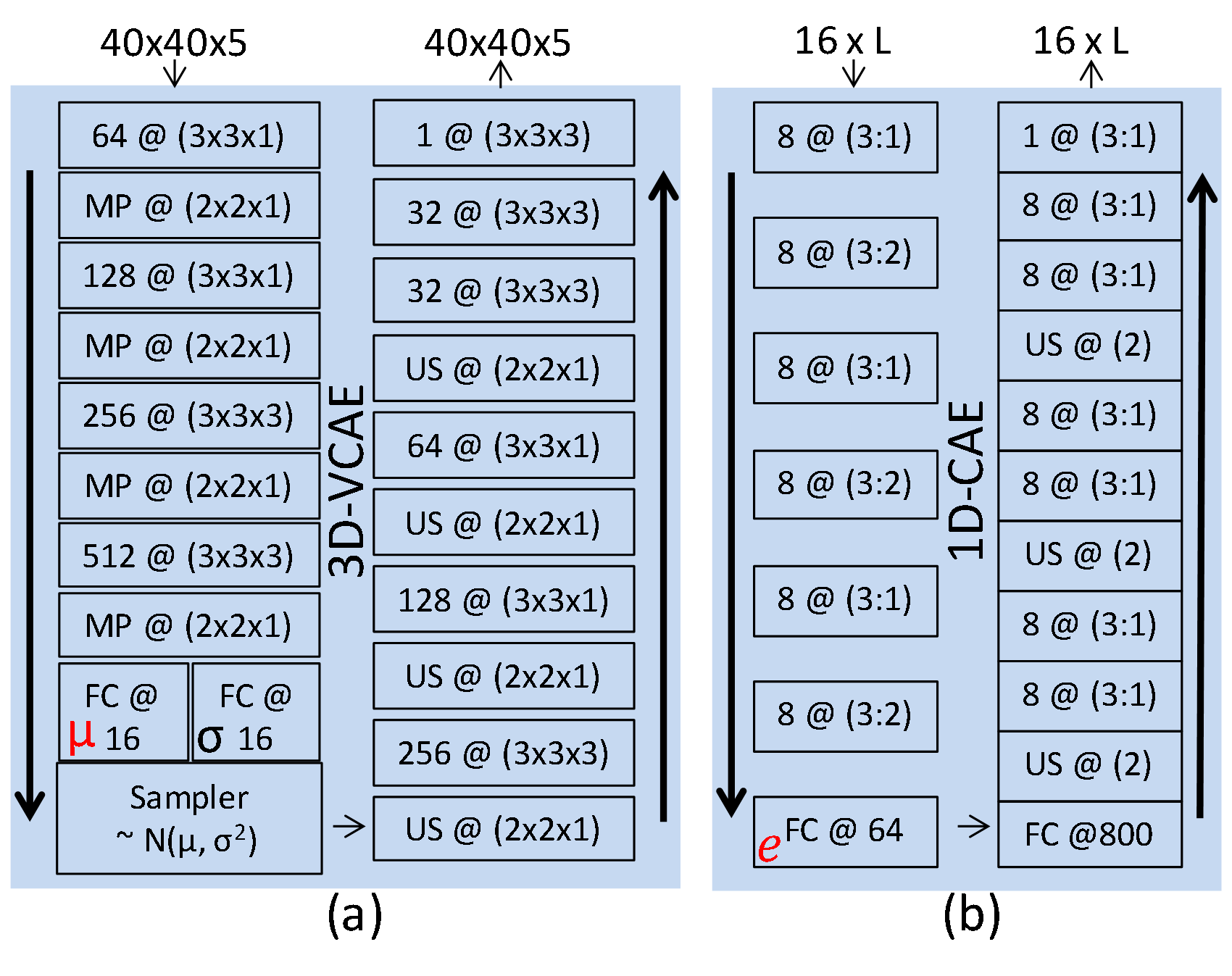}
	\caption{Architectures of autoencoders. (a) A 3D variational convolutional autoencoder (3D-VCAE), its input and outputs are volumes of 40x40x5 voxels, where the input is encoded into an encoding of size 16.
	Key: $N_{kernel}$@$size_{kernel}$ is a convolutional layer with $N_{kernel}$ kernels of size $size_{kernel}$. MP@$size_{kernel}$ is a max-pooling layer with kernel size $size_{kernel}$. US@$size_{kernel}$ is an upsampling layer with kernel size $size_{kernel}$. FC@$N_{units}$ is a fully connected layer with $N_{units}$ units. Once the 3D-VCAE is trained, the output of the $\mu$ layer is used to generate encodings for the input. (b) A 1D convolutional autoencoder (1D-CAE), its input and outputs are 16xL sequences of encodings. Each sequence is padded into a maximal length of 800, and is encoded into an encoding of size 64. Key: $N_{kernel}$@($size_{kernel}$:$size_{stride}$) is a 1D convolutional layer with $N_{kernel}$ kernels of size $size_{kernel}$ and stride of $size_{stride}$. US@$size_{kernel}$ is a 1D upsampling layer with kernel size $size_{kernel}$. The 1D-CAE is applied separately, but with shared weights, to each of the 16 1D-sequences. Once the 1D-CAE is trained, the output of the $e$ layer is used to generate encodings for each input sequence.
   Descending and ascending arrows represent the encoder and the decoder in each autoencoder, respectively.}
	\label{fig:archs}

\end{figure}

 VAEs are generative models, which approximate data generating distributions \cite{kingma2013auto}. Through approximation and compression, the resulting models have been shown to capture the underlying data manifold; a constrained, smooth, continuous, lower dimensional latent (feature) space where data is distributed \cite{kingma2014semi,goodfellow2016deep}. 
 \add{Having in mind possible reconstruction errors of the encoding sequences by the second CAE, a \textit{variational} CAE is chosen as the first CAE for the ability of its decoder to handle small variations in the encodings \cite{kingma2014semi}}. Inspired by these advantageous properties of the latent space, a VCAE is employed to compress and encode local volumes along the artery.  
 To capture local volumetric characteristics of the artery, the input to the 3D-VCAE is set to a volume of 40x40x5 voxels, centered around a coronary artery centerline point. The size of the input is chosen so that it contains the whole arterial lumen and the vicinity of the artery \cite{cury2016cad}. The output of the encoder in the 3D-VCAE is set to 16; i.e. an encoding in a $R^{16}$ latent space. \add{The dimension of the input volume and encoding size are determined in preliminary experiments to balance between the compactness (i.e. size of the encoding) and the expressiveness of the encodings (i.e. reconstruction error). Table \ref{tab:comp_hyper} lists these findings}. To encode the complete artery, overlapping volumes with stride of 1 are extracted and encoded with 3D-VCAE (Fig.~\ref{fig:encoding_flow}). This results in 16xL encodings, where L is the length of the artery. \add{To reconstruct a complete artery, the middle slices of each overlapping reconstructed volume are used.} The 3D-VCAE architecture used in this work is shown in Fig.~\ref{fig:archs}(a). In the 3D-VCAE , batch normalization \cite{Ioff15} layers and rectified linear units (ReLUs) are used after all convolution layers except the encoder and decoder output layers.

\begin{table*}[]
	\begin{tabular}{@{}lllllll@{}}
		\toprule
		\begin{tabular}[c]{@{}l@{}}Input size\\ of CAE 1\end{tabular} & \begin{tabular}[c]{@{}l@{}}Encoding size\\ of CAE 1\end{tabular} & \begin{tabular}[c]{@{}l@{}}Encoding size\\ of CAE 2\end{tabular} & \begin{tabular}[c]{@{}l@{}}Encoding\\strategy\end{tabular} &\begin{tabular}[c]{@{}l@{}}Reconstruction MAPE\\ (within lumen)\end{tabular} & \begin{tabular}[c]{@{}l@{}}Total encoding\\size\end{tabular} & \begin{tabular}[c]{@{}l@{}}Average \\ AUC\end{tabular} \\ \midrule
		40x40x5                                                              & 8                                                              & 64 & 1D                                                                    & $32.7\pm12.6$                                                                & 512                                                      & $0.74\pm0.03$                                          \\
		40x40x5                                                              & 16                                                             & 64&  1D                                                                    & $29.1\pm9.8$                                                                 & 1024                                                     & $0.81\pm0.02$                                          \\
		40x40x5                                                              & 32                                                             & 64   &   1D                                                                & $17.7\pm5.2$                                                                 & 2048                                                     & $0.63\pm0.02$                                          \\ \midrule
				\addminor{40x40x5}                                                              & \addminor{$16^*$}                                                                      & \addminor{$64^*$} &\addminor{PCA}                                                        & \addminor{$51.4\pm13.0$}                                                                 & \addminor{1024}                                                              & \addminor{$0.51\pm0.03$}                                          \\
				
		40x40x5                                                              & 16                                                                      & 64 &1D                                                        & $29.1\pm9.8$                                                                 & 1024                                                              & $0.81\pm0.02$                                          \\
		
		40x40x5                                                              & 16                                                                      & 1024 &2D                                                      & $40.7\pm13.6$                                                                & 1024                                                              & $0.68\pm0.03$                                          \\
		40x40x800                                                            & 1024                                                                    & - &Global                                                     & $79.0\pm18.1$                                                                & 1024                                                              & $0.52\pm0.01$                                          \\ \midrule
		40x40x5                                                              & 16                                                                      & 32  &    1D                                                       & $29.8\pm9.6$                                                                 & 512                                                      & $0.78\pm0.04$                                          \\
		40x40x5                                                              & 16                                                                      & 64   &    1D                                                      & $29.1\pm9.8$                                                                 & 1024                                                     & $0.81\pm0.02$                                          \\
		40x40x5                                                              & 16                                                                      & 128   &      1D                                                   & $23.6\pm8.4$                                                                 & 2048                                                     & $0.61\pm0.03$                                          \\ \midrule
		40x40x5                                                     & 16                                                                      & 64 &     1D                                                                & $29.1\pm9.8$                                                                 & 1024                                                              & $0.81\pm0.02$                                          \\
		40x40x13                                                    & 16                                                                      & 64  &     1D                                                               & $28.5\pm8.8$                                                                 & 1024                                                              & $0.80\pm0.03$                                          \\
		40x40x23                                                    & 16                                                                      & 64    &    1D                                                              & $32.1\pm10.5$                                                                & 1024                                                              & $0.72\pm0.04$                                          \\ \bottomrule
	\end{tabular}
	\caption{\add{Average reconstruction MAPE, (within lumen HU range), total size of the final encoding used for classification and the achieved AUC for artery-level classification across a range of different input sizes of first CAE (CAE 1), different encoding sizes of first and second CAE (CAE 2) or different encoding strategies: \addminor{PCA: using two consecutive principle component analyses}; 1D: using a 1D-CAE (as CAE 2); 2D: using a 2D-CAE (as CAE 2); Global: using a single 3D-VCAE applied to the complete artery}. \addminor{Please note that the proposed configuration (input size of 40x40x5 with 16/64 encoding sizes) is listed multiple times for easy comparison.$^*$Number of principle components.} }
	\label{tab:comp_hyper}
\end{table*}

\subsubsection{Sequential encoding with 1D convolutional autoencoder}
 When representing the coronary artery to determine the functionally significant stenosis
 according to FFR, characteristics along the artery, starting from the ostium to the most distal part of the artery, need to be taken into account \cite{Tayl13,Itu12,Nick15}. Therefore, to analyze the complete artery at once, the local encodings extracted previously by the 3D-VCAE along the length of the artery need to be merged.  
 To accomplish this, the feature map, consisting of L sets of 16 values of encoding generated by 3D-VAE at each coronary artery center point, is represented as L 1D sequences.  \add{As in the 3D-VCAE design, the size of the 1D-CAE encoding is determined in preliminary experiments to balance between the compactness (i.e. size of the encoding), the expressiveness of the encodings (i.e. reconstruction error) and the classification performance (AUC). Table \ref{tab:comp_hyper} lists these findings}. Each sequence consists of 1xL values, where L represents the length of the artery, i.e. number of coronary artery centerline points. To encode arteries with different lengths, sequences of encodings of short arteries were padded into a maximum length of 800, which corresponds to the \addminor{number of centerline points in the} longest artery in the dataset. This representation leads each sequence to represent a specific member of the encoding in the $R^{16}$  latent space along the artery (colored 1D signals in Fig.~\ref{fig:encoding_flow}). This, consequently, allows us to apply a 1D-CAE to each of the 16 sequences separately.
  The weights of the 16 1D-CAEs are shared, where each 1D-CAE encodes one of the 16 sequences into an encoding of a second latent space of 64 dimensions ($R^{64}$). This results in 1024 (16x64) features that represent the complete artery. The 1D-CAE architecture used in this work is shown in Fig.~\ref{fig:archs}(b). In the 1D-CAE, the exponential linear units (ELUs) are used after all convolutions layers except the encoder and decoder output layers.

%\subsection{Classification according to the presence of functionally significant stenosis}
\subsection{Classification of arteries and patients}\label{method_class}

Based on the extracted encodings from the encoding stage, arteries are classified according to the \add{need of further invasive evaluation, in the form of ICA.} This was defined by the invasively measured FFR.
 \add{As the standard deviation of the differences in repeated FFR measurements can reach up to $5\%$ \cite{Ntal10,Berr13,Petr13b,John14}, especially  in the so called "\textit{gray-zone}" \cite{Petr13b}, where the measured FFR is between 0.75 and 0.85}, in our experiments, the threshold on FFR value was set to 0.9. \add{This results} in a positive class with $FFR\le0.9$ representing arteries \add{requiring ICA to establish the need of intervention}, and a negative class with $FFR>0.9$ representing absence of functionally significant stenosis, \add{where ICA is not necessary.} The classification is performed using an SVM classifier with a linear kernel and an $L_1$ regularization. For each classified artery, the continuous output of the trained SVM is used to assign a predicted class.

As patients with suspected obstructive CAD undergo ICA to measure the FFR in all diseased coronary arteries, in this work, classification of patients is also performed. 
To classify patients, the highest output value of all classified arteries in a patient is used to assign a predicted class to the patient. The minimal FFR across the arteries of a patient is taken as a reference.

Classification performance is evaluated using a receiver operating characteristic (ROC) curve and the corresponding area under the ROC curve (AUC).

\section{Experiments and Results} \label{results}

\subsection{Encoding the artery}\label{encoding}

To train the 3D-VCAE and the 1D-CAE, a set of CCTA images of 50 patients, who did not undergo ICA and hence had no FFR measurements, were used. From these, 38 CCTA images were randomly selected for training, and the remaining 12 images were used for validation. In both sets, MPR volumes of the extracted arteries were used to train and validate the autoencoders. Both autoencoders' hyperparameters were determined in preliminary experiments using the validation set. \add{Please note that \addminor{no} data augmentation was performed for training the autoencoders.}
\addminor{As a baseline reconstruction, principal component analysis (PCA) was employed in a similar way as the two disjoint autoencoders: A first PCA was applied to all 40x40x5 voxels volumes along the artery to reduce each into 16 principal components. Then, a second PCA was applied to all outputs of the first PCA to reduce the dimensionality of the outputs into 1024 components. Table \ref{tab:comp_hyper} lists the findings.}

\begin{figure*}[h]
	\centering
	\includegraphics[width=1.0\textwidth]{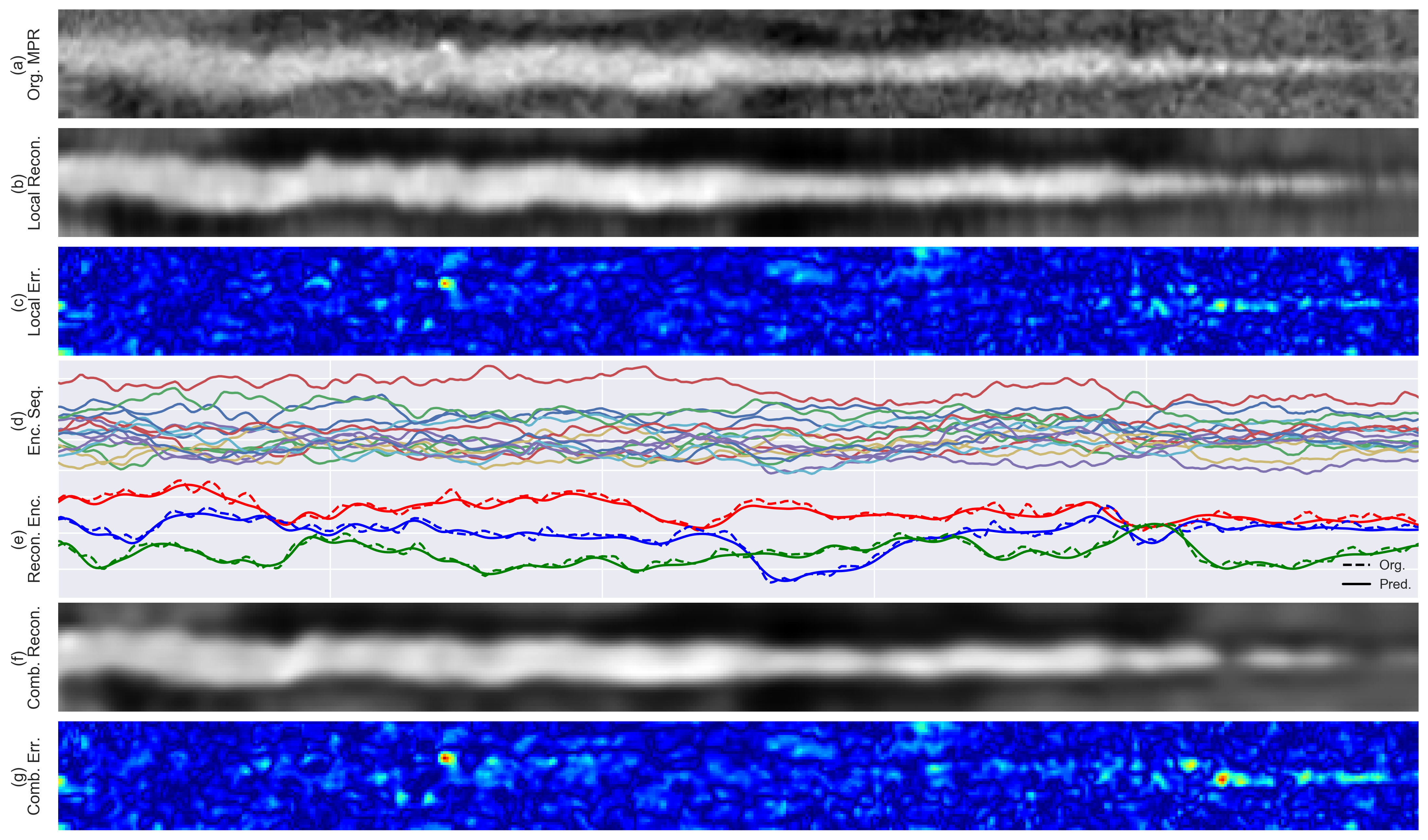}
	\caption{Examples of outputs from different stages in encoding and reconstructing a complete artery. (a) Original MPR of a complete artery; (b) The reconstructed MPR by only the 3D-VCAE. This was performed by extracting, encoding and reconstructing 40x40x5 voxels volumes around each point along the MPR centerline; (c) The absolute error between (a) and (b); (d) Encodings extracted by the 3D-VCAE, presented as continuous sequences running along the artery; (e) Three randomly selected encodings sequences of a complete artery (dashed line) which are encoded and reconstructed (solid line) with the 1D-CAE. (f) The reconstructed artery that was encoded and reconstructed back with the 3D-VCAE and 1D-CAE combined. (g) The absolute error between (a) and (f).
	}
	\label{fig:all_reconstructions}
	
\end{figure*}

To train and validate the 3D-VCAE, 40x40x5 voxels volumes were randomly extracted along centerlines of arteries in the training and validation sets, respectively. Mini-batches of 32 volumes were used to minimize the loss function with Adam optimizer \cite{kingma2014adam} with a learning rate $0.001$. The mean squared error between the input and the reconstructed volumes, and the Kullback-Leibler (KL) divergence with the reparameterization trick \cite{kingma2013auto} were employed as a loss function for the variational autoencoder. L2 regularization was used with $\gamma=0.001$ for all layers. Training was performed until convergence.
Fig.~\ref{fig:all_reconstructions}(a)-(b) show an example of a complete artery which was encoded and reconstructed with the trained 3D-VCAE. This was performed by extracting, encoding and reconstructing input volumes around each point along the MPR centerline. Fig.~\ref{fig:all_reconstructions}(c) shows the absolute reconstruction error, i.e. the absolute difference between the input and the reconstructed artery.
 Fig.~\ref{fig:all_reconstructions}(d) shows all 16 sets of encodings, presented as continuous sequences running along the artery. These sequences are to be encoded further in a later stage using the 1D-CAE into a fixed number of encodings.

To train and validate the 1D-CAE, arteries with the corresponding sets of 16 encodings sequences, obtained by the 3D-VCAE, were randomly chosen from the training and validation sets, respectively. Sequences of encodings of short arteries were padded into a maximum length 800, which corresponds to the longest artery in the dataset. Mini-batches of 32 sets of encodings sequences were used to minimize the loss function with Adam optimizer with a learning rate $0.001$. The masked mean squared error was employed as a loss function, where padded values in the input sequences did not contribute to the loss value or its gradients and were therefore ignored. L2 regularization was used with $\gamma=0.001$ for all layers. Training was performed until convergence. Fig.~\ref{fig:all_reconstructions}(e) shows an example of 3 randomly chosen encoding sequences of a complete artery which were encoded and reconstructed with the trained 1D-CAE.

\begin{figure}[h]

		\centering
		\includegraphics[width=1\linewidth]{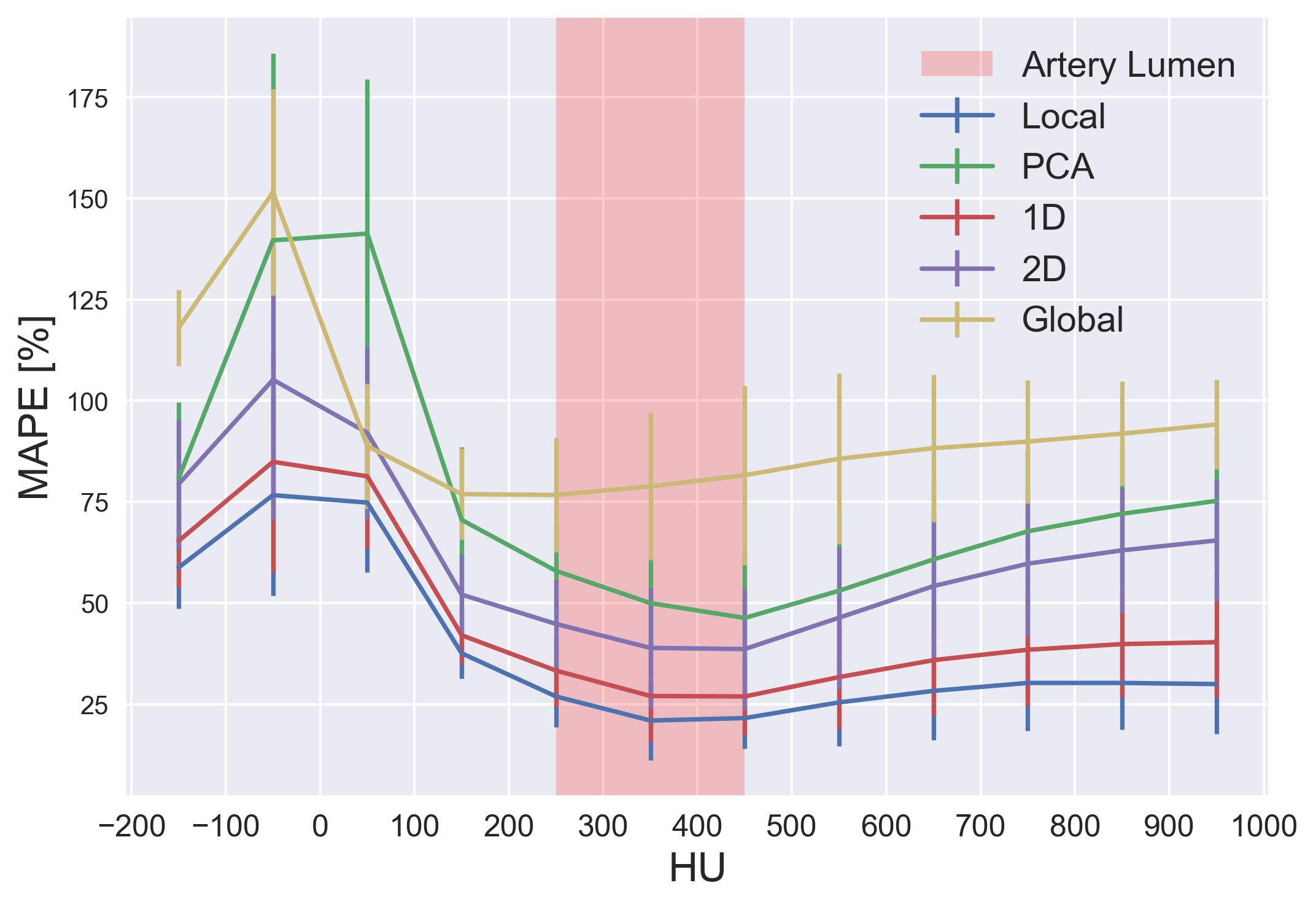}

	\caption{The average and standard deviation of the mean absolute reconstruction percentage errors (MAPE) obtained by the different reconstruction strategies, across a range of CT Hounsfield units (HU). Artery lumen: indicates the typical range (250-450 HU) of the CT values of the coronary artery lumen. Local: local reconstructions by the 3D-VCAE only (as in Fig.~\ref{fig:all_reconstructions}(b)); \addminor{PCA: baseline reconstruction strategy, using 2 consecutive PCAs}; 1D: combined reconstruction strategy, using the 1D-CAE (as in Fig.~\ref{fig:all_reconstructions}(f)); 2D: combined reconstruction strategy, using the 2D-CAE (as in Fig.~\ref{fig:diff_reconstr_compared}(d)); Global: global reconstruction strategy, using a 3D-VCAE (refer to Fig.~\ref{fig:global_vcae}) applied to the complete artery (as in Fig.~\ref{fig:diff_reconstr_compared}(f)).
		 }
	\label{fig:maes}
	
\end{figure}

To demonstrate the effectiveness of the proposed combined two-stage encoding approach in preserving the original shape and appearance of the artery, both disjoint trained autoencoders were combined and tested on complete arteries. To accomplish this, an inference with four steps was performed. First, the encoder of the 3D-VCAE was applied to local volumes along the MPR volume of a complete artery, resulting in 16 sequences of encodings with L values each. Second, the encoder of the 1D-CAE encoded the sequences into a single encoding vector of 1024 values. Third, the decoder of the 1D-CAE decoded the encoding vector back to 16 encodings sequences. Last, the decoder of the 3D-VCAE reconstructed those reconstructed encodings sequences to the original MPR volume size. Fig.~\ref{fig:all_reconstructions}(a),(f),(g) show an example of a complete artery that was encoded with the combined strategy, reconstructed back to the original volume dimensions, and the corresponding reconstruction error. Fig.~\ref{fig:maes} compares the average  mean absolute reconstruction percentage errors (MAPE) between the local reconstructions made by only the 3D-VCAE, \addminor{the baseline PCA reconstruction approach and} the combined approach reconstruction, across a range of CT Hounsfield units (HU). \add{As the MAPE might be misleading around small image values or image values equal to zero, the range of intensity values characteristic for coronary artery lumen (250-450 HU) is highlighted.} Fig.~\ref{fig:all_reconstructions} and Fig.~\ref{fig:maes} demonstrate the high resemblance and the low reconstruction error between the results of the local and the combined approaches compared to the original volume.

\subsection{Evaluation of alternative encoding strategies}\label{others}

To demonstrate that the proposed combined encoding strategy is advantageous, two additional encoding strategies were evaluated and compared to the proposed sequential disjoint autoencoders.

\begin{figure}[h]

	\centering
	\includegraphics[width=0.5\linewidth]{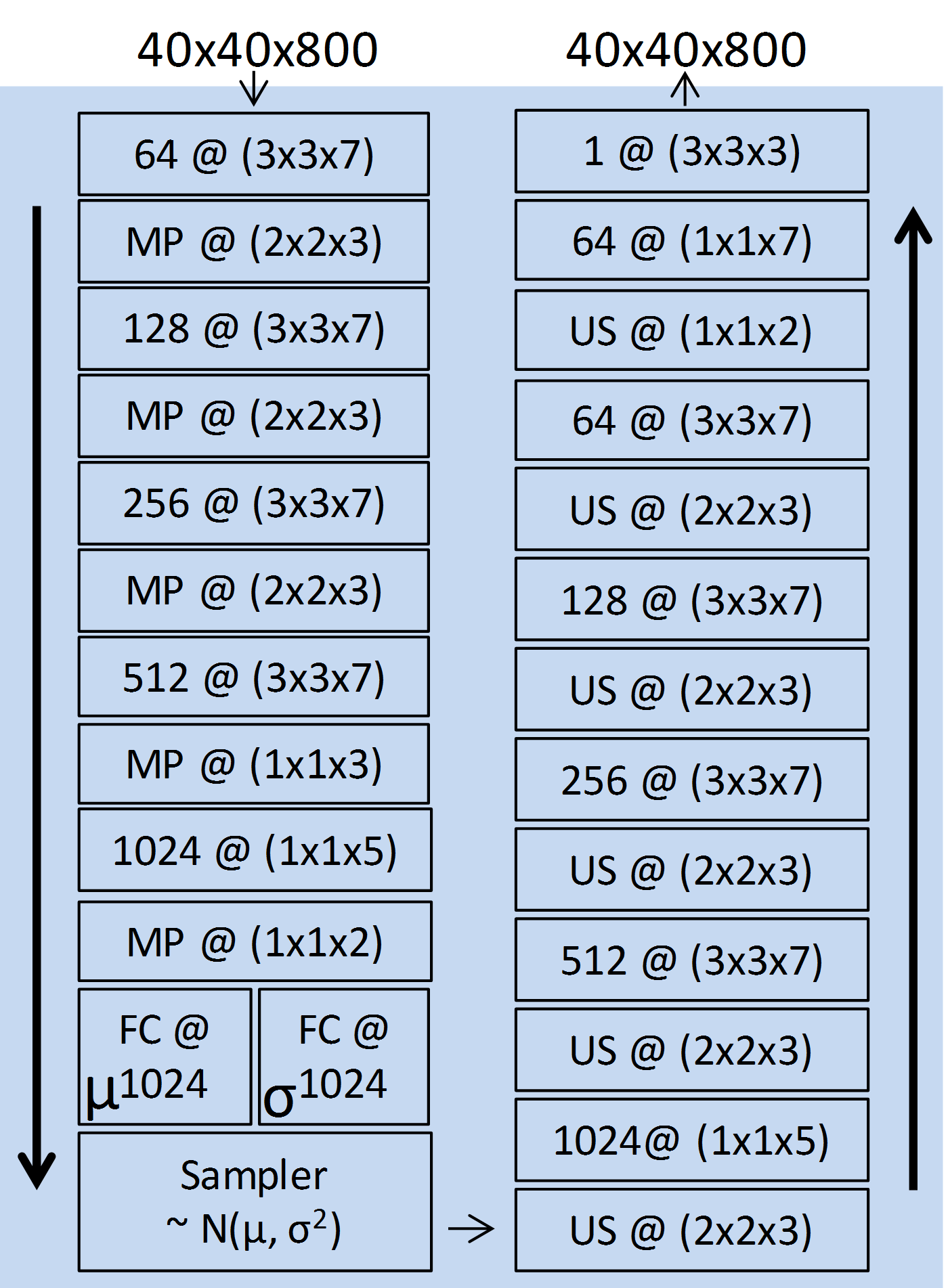}

	\caption{ The 3D-VCAE architecture used to encode a complete artery, where its input and outputs are MPR volumes of a complete artery of size 40x40x800 voxels. Keys are the same as in Fig.~\ref{fig:archs}.
	}
	\label{fig:global_vcae}
	
\end{figure}

First, the most straightforward approach was evaluated, where a single autoencoder analyzes the complete MPR volume, encodes it to a fixed number of encodings (1024), and reconstructs it back to the input size. To handle arteries with different lengths, shorter arteries were padded to the maximal artery length in our dataset (800). Hence, the input of the autoencoder was defined as 40x40x800. The architecture of the evaluated VCAE is shown in Fig.~\ref{fig:global_vcae}.

Second, the 1D-CAE used in the combined encoding strategy (Fig.~\ref{fig:encoding_flow}) was replaced by a 2D-CAE to jointly process the encoding sequences. While the proposed 1D-CAE \add{encoded each sequence of encodings} separately, the here evaluated 2D-CAE \add{mutually} encoded all sequences of encodings. This was performed by representing the encodings map as a 2D image and applying two-dimensional convolutional kernels. The architecture of the evaluated 2D-CAE is identical to the 1D-CAE (Fig.~\ref{fig:archs}(b)), but the convolutions were performed by applying two-dimensional kernels of the same size (3x3). Additionally, dropout of 0.1 was applied between fully connected layers to avoid overfitting.

\begin{figure*}[h]

	\centering
	\includegraphics[width=1\linewidth]{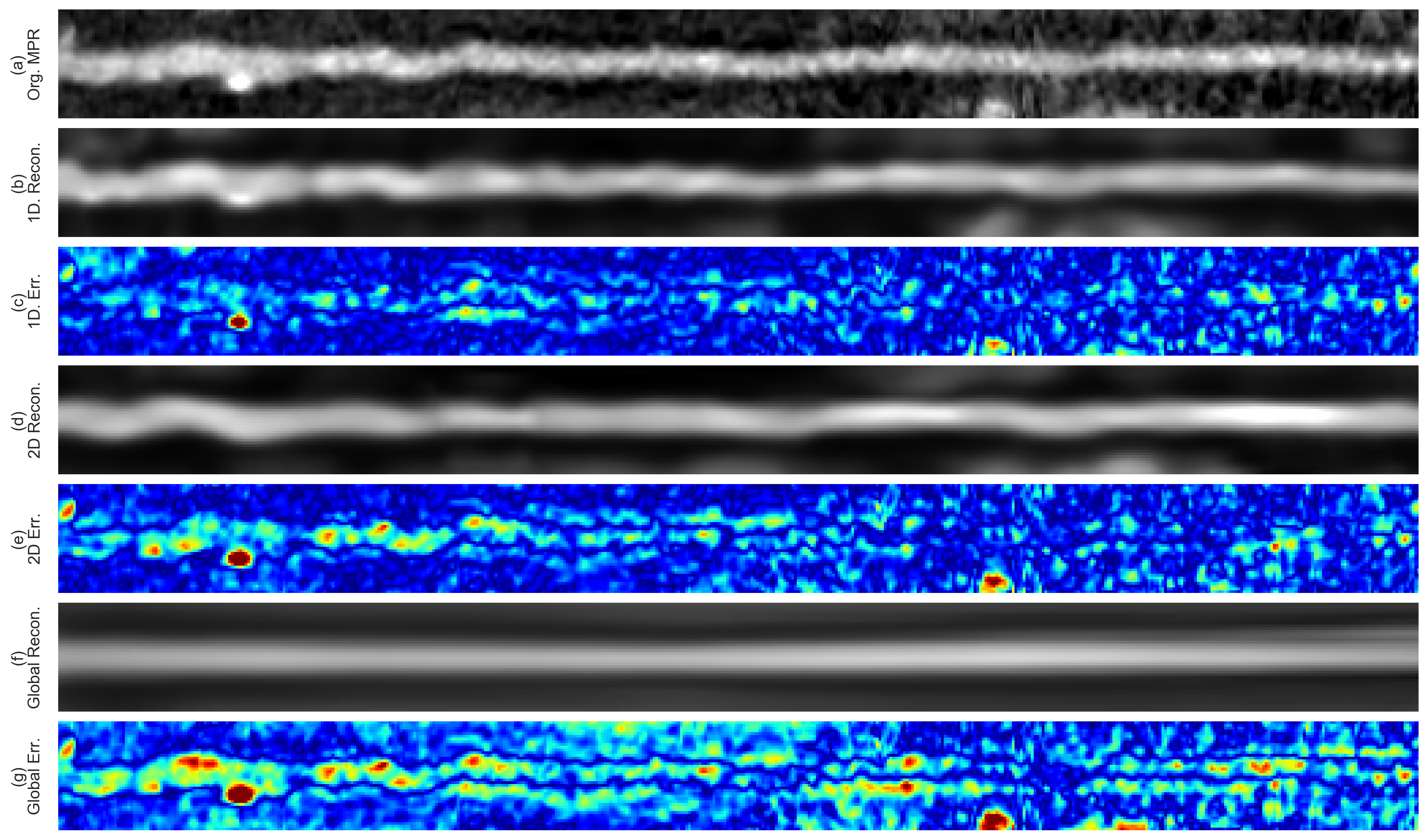}

	\caption{ 
		Examples of different reconstructions of a complete artery by different reconstruction strategies. (a) Original MPR of a complete artery; (b) The reconstructed MPR by the proposed combined 3D-VCAE and 1D-CAE; (c) The absolute error between (a) and (b);  (d) The reconstructed MPR by the combined 3D-VCAE and 2D-CAE; (e) The absolute error between (a) and (d); (f) The reconstructed MPR by the global 3D-VCAE (refer to Fig.~\ref{fig:global_vcae}), applied to the complete artery; (g) The absolute error between (a) and (f).
	}
	\label{fig:diff_reconstr_compared}
	
\end{figure*}

The two additional autoencoders were trained and validated in a similar manner as in the combined encoding strategy. In the case of the 2D-CAE, the trained decoder of the 3D-VCAE was used to reconstruct the MPR volume. Fig.~\ref{fig:diff_reconstr_compared} shows an example of an artery that was encoded and reconstructed using the two evaluated auto-encoding strategies, and compared with the reconstruction of the proposed combined encoding strategy. Fig.~\ref{fig:maes} also shows the average MAPE across a range of different HUs. Both figures demonstrate a clear advantage for the reconstruction of the combined strategy over the two additionally evaluated approaches.

\subsection{Classification of arteries and patients}\label{class_exp}

Classification of arteries was performed using the arteries' encodings, extracted by the two disjoint autoencoders (Section \ref{encoding}), \add{and an SVM classifier. In preliminary experiments, different classifiers (logistic regression, random forest) and various SVM configurations, including different regularizations and kernels types, were tested. The best performance was achieved using \addminor{a} linear $L_1$-regularized SVM}. All 50 CCTA images of patients used in training and validation of the autoencoders were excluded. Thus, CCTA images of 137 patients and the corresponding reference FFR measurements in 192 different arteries were used for this analysis.
To assess the performance and the robustness of the classification, 1000 stratified Monte-Carlo cross-validation experiments were performed. \add{In each experiment, 10 arteries were used as a test set, and the remaining arteries were used as a training set. The assignment of arteries to test or training sets was random, however it insured that arteries from the same patient were included either in the training or the test set.} Optimal SVM parameters were selected in every experiment using a grid search on the training set only.

\begin{figure}[h!]
	\centering
	\subfloat[\label{roc_vessel}]{%
				\includegraphics[width=1\linewidth]{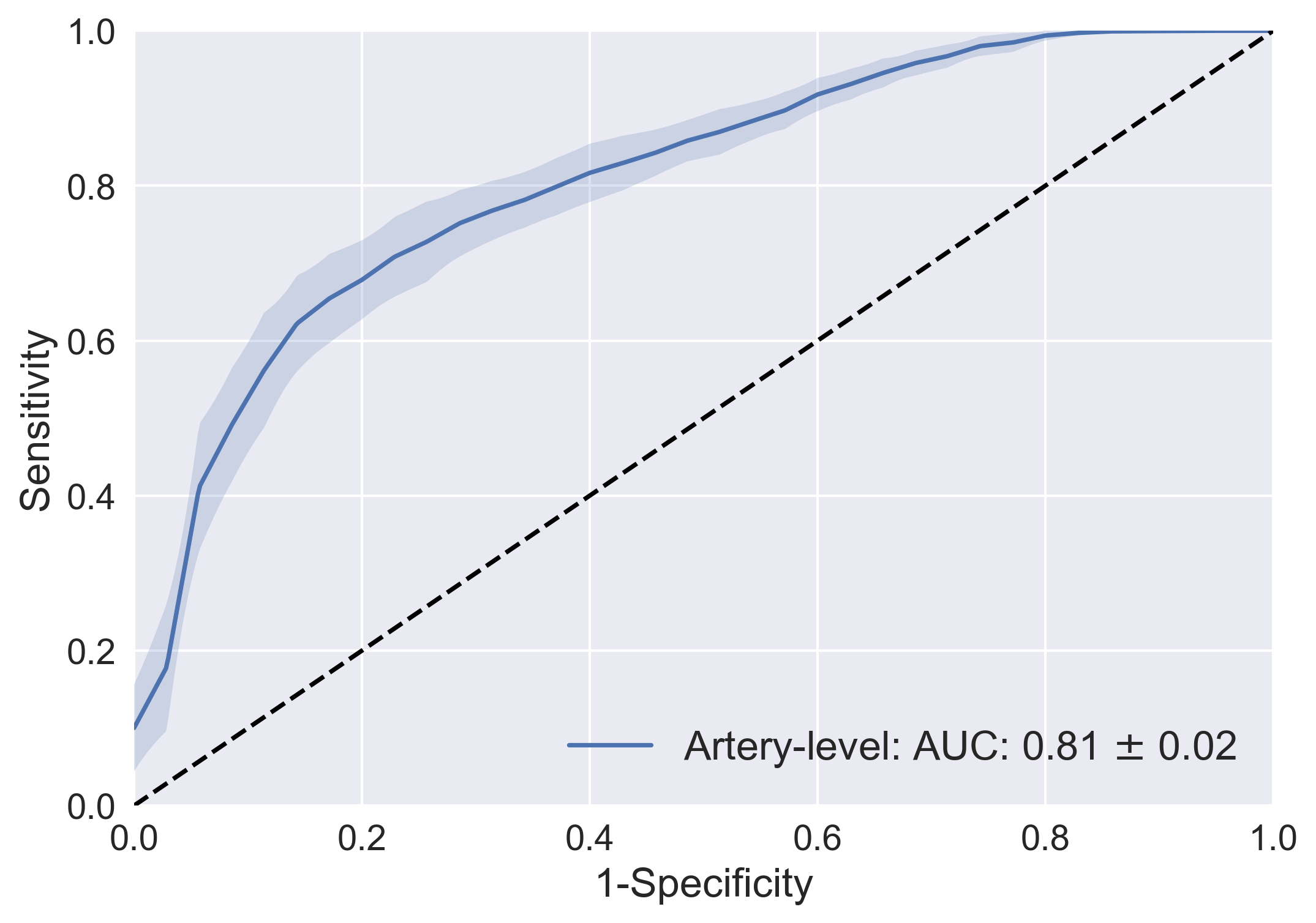}}
	\hfill
	\subfloat[\label{roc_patient}]{%
		\includegraphics[width=1\linewidth]{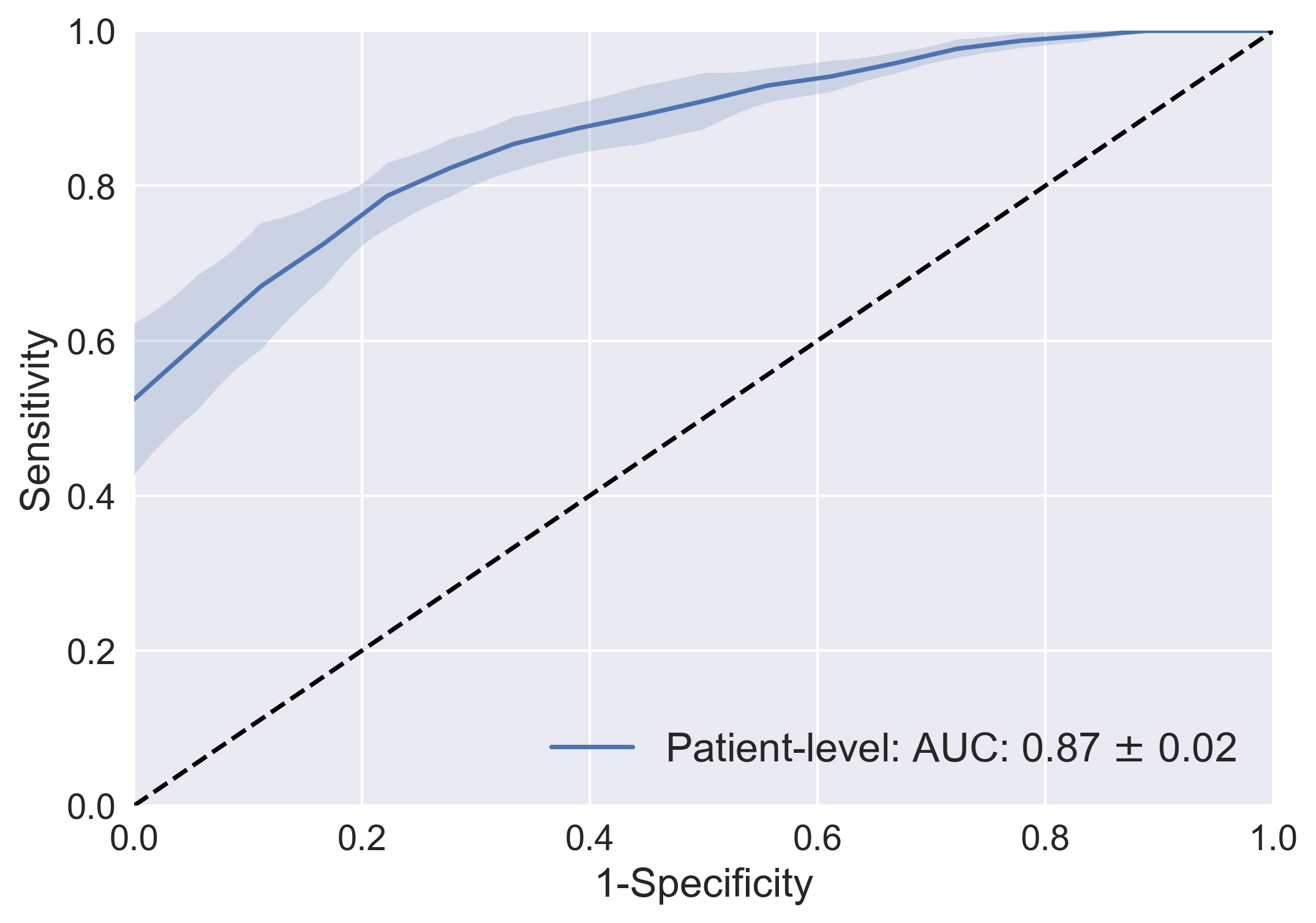}}

	\caption{Average ROC curves, and corresponding area under curve (AUC), for classification of (a) arteries and (b) patients \add{requiring ICA}. The classification was performed using the extracted encodings of the arteries and a support vector machine classifier. The shaded area represents the standard deviation of the sensitivity across the cross validation experiments.}
	\label{fig:cv_roc}
\end{figure}

The obtained results are shown in Fig.~\ref{fig:cv_roc}. On the artery-level, an average AUC of $0.81\pm0.02$ was achieved, while on the patient-level, an average AUC of $0.87\pm0.02$ was achieved. \add{Table \ref{tab:perf_ffr} lists the average diagnostic accuracy on the artery- and patient-levels across four different ranges of FFR measurements, and Table \ref{tab:acc_artery_label} lists the achieved performance in the three main coronary arteries. Moreover, implemented in Keras with TensorFlow, the runtime of encoding and classifying a single MPR volume of a complete coronary artery was on average 11 seconds, while using a single NVIDIA TITAN X (\addminor{P}ascal) GPU with an Intel Xeon machine with 256 GB RAM.}

% \usepackage{booktabs}
\begin{table}
\resizebox{\linewidth}{!}{
\centering

	\begin{tabular}{@{}lcccc@{}}
		FFR Range        & \begin{tabular}[c]{@{}c@{}}N\\ Arteries\end{tabular}& \begin{tabular}[c]{@{}c@{}}Artery-level\\ Accuracy\end{tabular} &\begin{tabular}[c]{@{}c@{}}N\\ Patients\end{tabular}& \begin{tabular}[c]{@{}c@{}}Patient-level\\ Accuracy\end{tabular} \\ \midrule
		$FFR\leq0.7 $   &\addminor{32} & 0.66                                                            &\addminor{26} & 0.70                                                             \\ \midrule
		$0.7<FFR\leq0.8$& \addminor{52}& 0.75                                                            &\addminor{41} & 0.76                                                             \\ \midrule
		$0.8<FFR\leq0.9$& \addminor{73}& 0.79                                                            & \addminor{52}& 0.78                                                             \\ \midrule
		$FFR\geq0.9$    & \addminor{35}& 0.73                                                            &\addminor{18} & 0.80                                                             \\ \bottomrule
	\end{tabular}
	\caption{\add{Average diagnostic accuracy for the detection of arteries and patients requiring ICA on the artery- and patient-levels \addminor{shown in four different subgroups corresponding to four ranges of FFR measurements.}} \addminor{N indicates the number of arteries and patients in each subgroup.}}
	\label{tab:perf_ffr}
}
\end{table}

\begin{table}[]
	\begin{tabular}{lcc}

	Artery & N   & Accuracy \\ \hline
	LCX    & 52  & 0.65     \\ \hline
	RCA    & 36  & 0.71     \\ \hline
	LAD    & 104 & 0.80     \\ \hline
\end{tabular}
\caption{\add{Average diagnostic accuracy for the detection of arteries requiring ICA \addminor{shown in three subgroups, corresponding to} the three main coronary arteries: Left circumflex artery (LCX), right coronary artery (RCA) and left anterior descending (LAD). N indicates the number of arteries in the data set.}}
\label{tab:acc_artery_label}
\end{table}

\subsection{\addminor{Evaluation of alternative classification strategies}}\label{other_class_results}

\addminor{
To demonstrate the effectiveness of the proposed classification scheme, we have performed several additional classification experiments that can be divided into two categories. 
}

\addminor{
First, the influence of the FFR threshold was investigated. As in Section \ref{method_class}, arteries were classified into positive or negative class with a binary SVM classifier using the extracted encodings. Different thresholds on reference FFR values were applied, resulting in different class interpretations: When a threshold of 0.7 was applied, positive classes represented arteries in need for an invasive intervention without the need of establishing the FFR value first. When a 0.8 threshold was applied, positive class represented arteries with a functionally significant stenosis. 
}

\addminor{
Second, the influence of regression vs. classification was investigated. In contrast to the former classification scenarios where retraining the SVM classifier was needed for each different FFR threshold, here, a single SVM \textit{regressor} was trained to estimate continuous values of FFR (i.e. regression) and then an FFR threshold (0,7, 0.8 or 0.9) was applied to output the binary classes. 
}

\begin{figure}
	\centering
	\includegraphics[width=1.\linewidth]{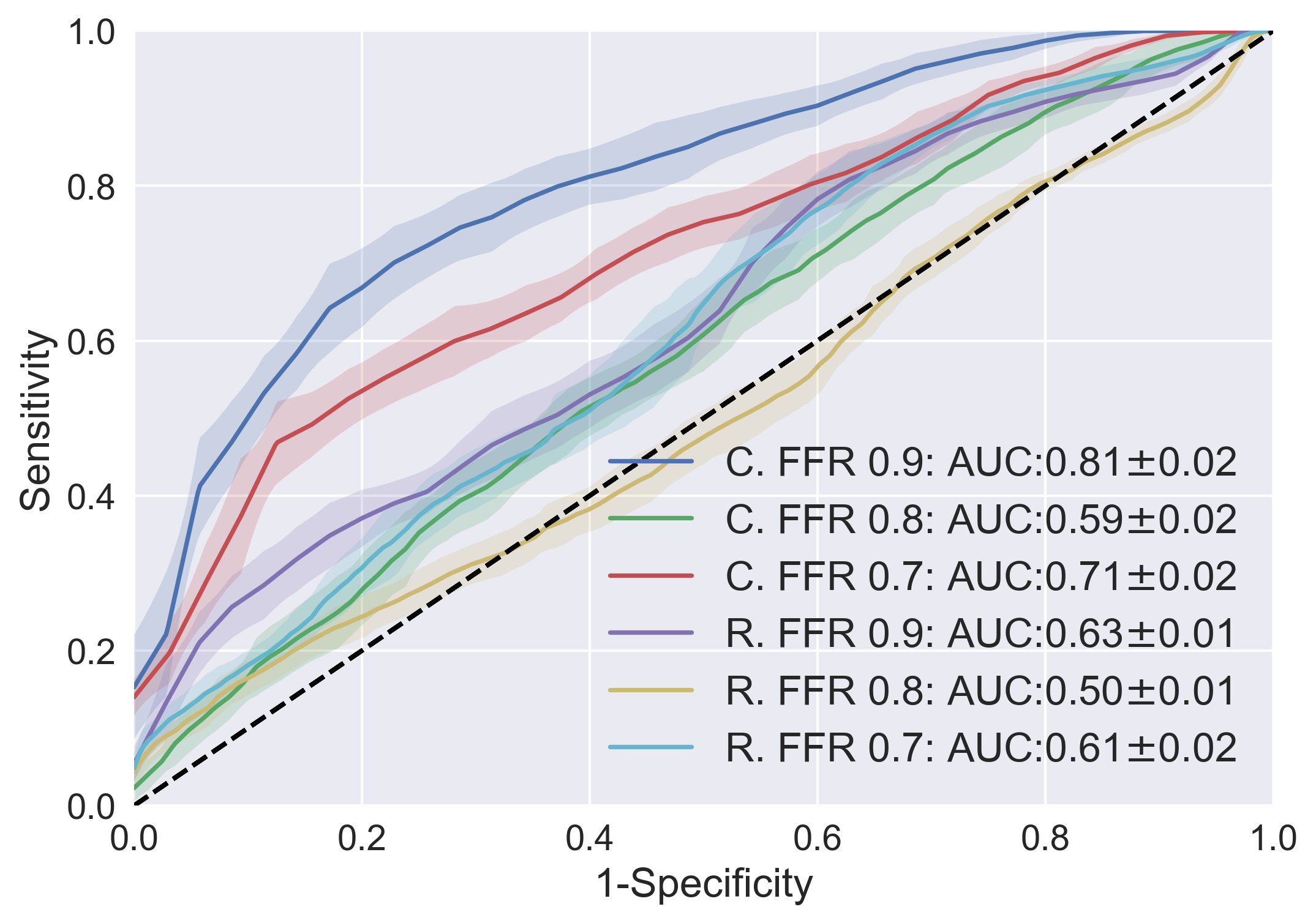}
	\caption{\addminor{Average ROC curves, and corresponding area under curve (AUC), for classification of arteries under different classification strategies and various FFR thresholds (0,7, 0.8, 0.9). The classification was performed using the extracted encodings of the arteries and a support vector machine classifier to: (1) binary classify arteries (C.) or (2) regress continuous FFR value and subsequently use the FFR threshold to produce the binary output (R.). The shaded areas represent the standard deviation of the sensitivity across the cross validation experiments.}}
	\label{fig:other_class_reg}
\end{figure}

\addminor{
The ROC curves showing the results are given in Fig.~\ref{fig:other_class_reg}. The results show that binary classifications outperform the corresponding regression experiments, regardless of FFR threshold. Moreover, when an FFR threshold of 0.7 was used, performance of the both classification and regression approaches were moderate, while the experiments using threshold of 0.8 on FFR values showed lowest areas under the ROC curves.
}

\subsection{Comparison with other FFR classification methods}\label{others}

\add{We compare our classification performance with the reported results of previous methods. These methods either analyzed the blood flow in the coronary arteries \cite{Norg14,Coen15,coenen2018diagnostic}, requiring a highly accurate segmentation of the arterial lumen, or analyzed the LV myocardium \cite{Han17,zreik2018deep}, requiring a segmentation of the LV myocardium. Table \ref{list:other_methods_results} lists the results as originally reported. However, all of the compared methods were evaluated on different datasets that included different patients cohorts. Moreover, the compared methods employed an FFR cut-off value of 0.8 to define the functional significance of a stenosis, while in this study, a 0.9 cut-off point was used. Therefore, these results only indicate the differences in approaches and should not be directly compared.}

\begin{table*}
	\centering
	\caption{Performance comparison with previous work. Table lists number of evaluated patients and arteries, achieved accuracy and the area under the ROC curve (AUC) per-patient and per-artery for classification according to FFR as reported in the original studies. Please note that these methods use different FFR thresholds and perform different analyses: either analyzing of the blood flow in the coronary arteries (\textit{Flow}), detecting ischemic changes directly in LV myocardium (\textit{Myo.}), or, as proposed in this work; classifying coronary arteries with features extracted by CAEs. }
	\label{list:other_methods_results}

	\resizebox{\textwidth}{!}{%
		\begin{tabular}{clcccccccl}
			\hline

			& \multicolumn{1}{l}{} &                                           &                                          & & \multicolumn{2}{c}{\textbf{Per-patient}}                                 & \multicolumn{2}{c}{\textbf{Per-artery}}

			\\

			\cmidrule(lr){6-7}
			\cmidrule(lr){8-9}
			
			& \textbf{Study}       & \multicolumn{1}{l}{\textbf{Patients}} & \multicolumn{1}{l}{\textbf{Arteries}}& \multicolumn{1}{l}{\textbf{\add{$FFR\leq$}}} & \multicolumn{1}{l}{\textbf{Accuracy}} & \multicolumn{1}{l}{\textbf{AUC}} & \textbf{Accuracy} & \multicolumn{1}{l}{\textbf{AUC}} & \multicolumn{1}{l}{\textbf{Requirements}} \\ \hline
			\multirow{3}{*}{\rotatebox[origin=c]{90}{Artery}} 
			%& \cite{Min12}         & 252                                       & -                                        & 0.71                                  & 0.81                             & -                 & -                                \\ \cline{2-8} 
			& N{\o}rgaard et al.\cite{Norg14}        & 254                                       & 484   &         0.8                          & 0.81                                  & 0.90                             & 0.86              & 0.93     & \add{Artery lumen segmentation}                        \\ \cline{2-10} 
			%& \cite{Renk14}        & 53                                        & 67                                       & 0.86                                  & -                                & 0.85              & -                                \\ \cline{2-8} 
			& Coenen at al.\cite{Coen15}        & 106                                       & 189       &            0.8                   & -                                     & -                                & 0.74              & - & \add{Artery lumen segmentation}

			\\ \cline{2-10} 
			& Coenen at al.\cite{coenen2018diagnostic}        & \add{303}                                       & \add{525}     &               0.8                  & \add{0.71}                                     & -                                & \add{0.78}              & \add{0.84} 	& \add{Artery lumen segmentation}

			\\ \Xhline{2pt}
			
			\multirow{2}{*}{\rotatebox[origin=c]{90}{Myo.}} 
			
			& Zreik et al.\cite{zreik2018deep}                 & 126                                       & -     &                          0.8         & 0.64                                  & 0.66                             & -                 & - & \add{LV myocardium segmentation}
			\\ \cline{2-10} 
			
			& Han et al.\cite{Han17}         & 252                                       & 407   &           0.8                        & 0.63                                  & -                                & 0.57              & -   & \add{LV myocardium segmentation}

			\\ \Xhline{2pt}
			
			\multirow{2}{*}{\rotatebox[origin=c]{90}{}} & Proposed         & 137                                       & 192    &          0.9                        & 0.80                                  & 0.87                                & 0.78              & 0.81              & \add{Artery centerline tracking}

		\end{tabular}%
	}
\end{table*}

\section{Discussion}\label{discussion}
	
A method for automatic and non-invasive \add{identification of patients requiring evaluation with invasive coronary angiography has been presented.} The method \add{\addminor{analyzes} complete coronary arteries} with two convolutional autoencoders that characterize the MPR volume of each artery with general robust features, and encode the complete artery into a fixed number of encodings to reduce the dimensions of the input. Then, these encodings are used with an SVM classifier to identify arteries with functionally significant stenosis in an supervised manner. Unlike previous methods that detect functionally significant stenosis by relying either on the coronary artery lumen segmentation \cite{Tayl13,Itu12,Nick15,itu2016machine} or the left ventricle myocardium segmentation \cite{zreik2018deep, Xion15}, the proposed method requires only the coronary artery centerline as an input along with the CCTA scan. Artery centerline extraction is a simplified task compared to myocardium segmentation and to the arterial lumen segmentation, where the latter occasionally requires substantial manual interaction, especially in diseased population with heavily calcified arteries. In this work, to extract the coronary artery centerlines, we have employed our previously designed method for artery centerline extraction \cite{wolterink2019coronary}. However, any other manual, semi-automatic or automatic method could be employed instead.

As the dimensions of MPR volumes of complete arteries are large, and the reference labels are only provided on the artery level, employing a straight-forward supervised 3D-CNN to detect the functional significance of a stenosis is far from feasible. Therefore, here, we have used unsupervised learning to characterize and encode each artery before employing a supervised classifier to detected abnormal FFR. To do so, two disjoint CAEs, that were applied sequentially, were employed. This is contrary to the more common approach of using a single CAE that encodes the complete artery volume at once. The results show that the learned encodings were able to represent the artery shape and appearance accurately, as was demonstrated qualitatively (Fig.~\ref{fig:all_reconstructions}) and quantitatively by the relatively small mean absolute error between the input and the reconstructed volumes (Fig.~\ref{fig:maes}).

The output of the combined encoding strategy was examined and compared to the output of the 3D-VCAE on local volumes. Fig.~\ref{fig:all_reconstructions} and Fig.~\ref{fig:maes} show that, in the range of CT values of the artery lumen (250-450 Hounsfield units), both the local and the combined encoders (with 1D-CAE) achieved satisfactory results, where the local approach was slightly advantageous. However, the lower error achieved with the local approach could be explained by the larger number of encodings used per artery compared to the combined approach. It can be noticed (Fig.~\ref{fig:all_reconstructions}(b) and (f)) that the reconstructions of both approaches preserved the shape and the morphology of the artery, while not being able to preserve the texture of neither the lumen nor the background. \add{Although the lumen of the coronary artery was accurately reconstructed by the proposed encoding method, some small calcifications within the artery were entirely or partially lost (Fig.~\ref{fig:all_reconstructions} and Fig.~\ref{fig:diff_reconstr_compared}, respectively). This might be due to the low number of encodings used in the 3D-VCAE (16). Beside increasing the size of the encoding, future work could address this by modifying the loss function of the 3D-VCAE to penalize such errors in reconstruction, or by over-sampling such calcifications in the training process.}

In the proposed combined encoding strategy, both autoencoders were disjoint during training and were combined only during inference, which could lead to error propagation. To overcome this, one alternative would be training both autoencoders simultaneously and  end-to-end. However, in preliminary experiments, this was proven difficult, mainly due to hardware limitations. Another alternative would be training the 3D-VCAE separately, and then, using its trained decoder during training the 1D-CAE. This could be done by directly minimizing the mean squared error between the original and the reconstructed MPR volumes of the complete artery instead of between its original and reconstructed encodings sequences. Such training might potentially compensate for errors or prevent error propagation between the two disjoint training processes. Future work might address this.

Additional encoding strategies were performed and compared with the proposed one. 
\add{Although other studies showed that a single 3D-CAE might be successfully used on large volumetric input \cite{myronenko20183d}, in \addminor{the proposed} work} training a single 3D-VCAE (Fig.~\ref{fig:global_vcae}), applied directly to the complete artery volume without a large reconstruction error, was proven infeasible (Fig.~\ref{fig:maes} and Fig.~\ref{fig:diff_reconstr_compared}). This could be due to \add{a number of reasons:} The very large number of trainable parameters ($\sim65\times10^6$) of the CAE, the high variability among the shapes and lengths of the arteries\add{, the atypical aspect ratio of the input (40x40x800), or the lack of large set of training data}. Moreover, treating all encoding sequences as a single 2D image and encoding this image with a 2D-CAE proved inferior when compared with the proposed 1D-CAE (Fig.~\ref{fig:maes} and Fig.~\ref{fig:diff_reconstr_compared}). This might be explained by the lack of local spatial relations between the different encodings \add{($\mu_0-\mu_{15}$\addminor{)} in Fig.~\ref{fig:encoding_flow} at a given location along the coronary artery}. These \add{local} spatial relations \add{among the encodings} motivate the use of \add{2D kernels in a typical 2D-CAE}, when analyzing natural or medical images.

The artery was represented by multiple sequences of encodings, obtained after applying the 3D-VCAE to local sub-volumes along the artery. To further encode the artery to a fixed number of encodings, a 1D-CAE was applied to each sequence of encodings separately. As the fully connected layer in 1D-CAE (layer $e$ in Fig.~\ref{fig:archs}(b)) expects a fixed number of inputs, the input sequences were padded to the maximal length of an artery in the dataset. \add{A masked loss function was employed during training the 1D-CAE to minimize the effect of such padding on the reconstruction error. Despite this masking, the proposed padding could affect the classification performance as a function of the artery length.}
To enable the autoencoder to handle variable length sequences, without the need of padding it, a recurrent autoencoder could be employed \cite{sutskever2014sequence}. In such a recurrent autoencoder, known as sequence-to-sequence autoencoders, a recurrent layer, with Gated Recurrent Units (GRUs) \cite{cho2014learning} or long short-term memory (LSTM) units \cite{hochreiter1997long}, replaces the fully connected layer in the proposed 1D-CAE, to recursively process and encode a sequential varying length input. Future work might investigate \add{such a recurrent autoencoder and its affect on the classification performance.}

Our experiments show that moderate classification performance on both the artery- and patient-levels was achieved, while using only features derived in an unsupervised manner from a CCTA of coronary arteries (Fig.~\ref{fig:cv_roc}). These results show that the proposed approach could potentially lead to a reduction in the number of patients that unnecessarily undergo invasive coronary angiography. For example, as seen in Fig.~\ref{fig:cv_roc}(b), at the sensitivity of 80\% or 90\% in detecting patients requiring ICA, i.e. those having $FFR\le0.9$, unnecessary ICA could have been prevented in 76\% or 53\% of the negative patients, i.e. those having $FFR>0.9$, respectively. 
\add{Moreover, we have compared the classification results across different ranges of FFR measurements (Table \ref{tab:perf_ffr}). The comparison shows slightly higher accuracies for $FFR>0.8$, on both the artery- and patient-levels. The reason might be that arteries with $FFR>0.8$ typically contain less plaque and therefore were better characterized by the autoencoders. A comparison of the diagnostic accuracy in the three main coronary arteries (Table \ref{tab:acc_artery_label}) shows that the highest accuracy was achieved for the LAD. This might be due the small number of available training examples for the LCX and RCA.}

Unlike this study, most previous methods that analyze blood flow \cite{Tayl13,Itu12,Nick15,itu2016machine} for detection of functionally significant stenosis as determined by invasive FFR \add{estimate continuous values of FFR along the coronary artery and localize} the functionally significant stenoses using a threshold of 0.8 on the determined FFR \cite{Petr13b}. However, in preliminary experiments \add{ for estimation of continuous FFR values (i.e. with regression) or applying} such a threshold, the proposed method \add{has not} achieved satisfactory results (\addminor{refer to Section \ref{other_class_results} and Fig.~\ref{fig:other_class_reg}}). The reason may be threefold. 	
\add{First, unlike other methods \cite{Tayl13,Itu12,Nick15,itu2016machine}, the here proposed method analyzes only a single coronary artery at once, while not taking into account the other arteries in the complete coronary artery tree. Analyzing the entire coronary tree might be crucial to differentiate between arteries with functionally significant or non-significant stenoses. Second, as the complete coronary artery is characterized as a whole using the CAEs, spatial information about a specific stenosis can not be retained. Third, the small number of encodings used in this work preserves the coarse shape and morphology of the analyzed artery (see Fig.~\ref{fig:all_reconstructions}(f)). However, fine morphology might be crucial for differentiating the functionally significant stenoses with FFR measurements in the most difficult range around the FFR of 0.8 \cite{Petr13b}.
Although these methods that analyze blood flow reported better results \cite{Tayl13,Itu12,Nick15,itu2016machine}, they are heavily dependent on the accuracy of coronary artery lumen segmentation \cite{tesche2017coronary}. Highly accurate lumen segmentation is extremely challenging task especially in patients with excessive atherosclerotic calcifications or imaging artefacts or stents \cite{Kiri13a}. As a result, these patients are typically not eligible for such analysis and are excluded\cite{lu2017noninvasive,Pontone2019,Sakuma2019}. In contrast, our method does not require lumen segmentation and therefore heavily diseased patients were not excluded. With a larger data set, future work may further investigate estimation of FFR on the continuous scale, possible performance enhancement when complete coronary artery tree is taken into analysis, and investigate different encoding approaches that may preserve fine morphology of the arteries.}

To conclude, this study presented an automatic and non-invasive analysis of the coronary arteries in CCTA for detection \add{of patients requiring invasive coronary angiography to establish the need of coronary intervention}.
The method is based on two disjoint convolutional autoencoders that characterize and encode volumes of complete coronary arteries into a set of encodings. Thereafter, a support vector machine classifier classifies arteries, employing these encodings, according to the presence of \add{abnormal} invasively measured FFR. The achieved moderate classification performance shows the feasibility of reducing the number of patients that unnecessarily undergo invasive FFR measurements.

%\section*{Acknowledgments}
	
%This study was financially supported by the project FSCAD, funded by the Netherlands Organization for Health Research and Development (ZonMw) in the framework of the research programme IMDI (Innovative Medical Devices Initiative); project 104003009.

	%\section{References}
	%\medskip

\bibliographystyle{IEEEtran}

%\bibliography{CAD}

\end{document}